\documentclass[12pt]{iopart}
\usepackage[utf8x]{inputenc}
\usepackage[T1]{fontenc}
\usepackage[english]{babel}
\usepackage{graphicx}
\usepackage{iopams}
\usepackage{algorithm}
\usepackage{algpseudocode}
\usepackage{url}

\begin{document}

\title{Finding network communities using modularity density}

\author{Federico Botta$^1$, Charo I. del Genio$^2$}
\address{$^1$ Centre for Complexity Science, University of Warwick, Coventry CV4 7AL, UK}
\address{$^2$ School of Life Sciences, University of Warwick, Coventry, CV4 7AL, UK}
\eads{\mailto{f.botta@warwick.ac.uk}, \mailto{C.I.del-Genio@warwick.ac.uk}}

\begin{abstract}
Many real-world complex networks exhibit a community
structure, in which the modules correspond to actual
functional units. Identifying these communities is a
key challenge for scientists. A common approach is to
search for the network partition that maximizes a quality
function. Here, we present a detailed analysis of a
recently proposed function, namely modularity density.
We show that it does not incur in the drawbacks suffered
by traditional modularity, and that it can identify
networks without ground-truth community structure, deriving
its analytical dependence on link density in generic
random graphs. In addition, we show that modularity
density allows an easy comparison between networks of
different sizes, and we also present some limitations that methods 
based on modularity density may suffer from. Finally, we introduce an efficient,
quadratic community detection algorithm based on modularity density
maximization, validating its accuracy against theoretical
predictions and on a set of benchmark networks.
\end{abstract}


\maketitle

\section{Introduction}
The last two decades have seen an explosion in the study
of complex systems, caused by the increasing relevance for
society of such large interconnected structures, and by
an unprecedented availability of data to analyze them. Many
of these systems can be modelled as networks, in which the
system elements are represented as nodes, and their interactions
as connections, or edges, linking them~\cite{Alb02,New03,Boc06}.
The network representation of complex systems has been used
in the social sciences~\cite{Was94,Sco00,New01,Laz09,Ves09},
in biology~\cite{Wil00,Jeo00,Tre12,Joh14}, and in studies
of technological systems~\cite{Alb99} and communication
systems~\cite{Sar15}. More recent work has focussed on the
multilayer nature of complex networks, introducing a new 
framework that is particularly useful for the analysis of
large complex data sets~\cite{Boc14,Kiv14}. Researchers
have applied complex systems techniques to a wide range
of disciplines, identifying and analyzing several defining
features of complex networks, such as the small world property~\cite{Mil67,deS78,Wat98,Ama00},
heterogeneous degree distributions~\cite{Bar99,del11}, clustering~\cite{New01_2,del13},
degree-degree correlations~\cite{Joh10,Wil14}, assortativity~\cite{New02},
synchronizability~\cite{del15}, and community structure~\cite{Gir02}.
Communities were originally studied in the context of social
networks, in which they are formed by groups of people that
share close friendship relations. However, communities of
densely connected modules have been observed in several
real-world and model networks of diverse nature~\cite{Pim79,Gar96,Fla02,Eri03,Kra03,Lus04,Gui05,Pal05,Are06,Res06,Hus07,Blo08,del11_2},
where, in general, they are defined as groups of nodes whose
internal connections are denser or stronger than those that
link nodes belonging to different groups. In all these cases,
the presence of communities directly influences the behaviour
of the system, where there is often a correspondence between
communities and functional units. Ever since the discovery
of community structure in real-world networks, a plethora
of techniques devoted to their detection has been introduced~\cite{Dan05,Lan08,For10,Muc10,Ste10,Dec11,Pei13,Tre15,New15}.
The challenge is both theoretical, in proposing a good mathematical
definition of what constitutes a community, and computational,
in developing good heuristics that can detect communities
in a reasonable time.

A common  way of investigating the community structure of networks
starts with the definition of a quality function, which assigns a
score to any network partition. Larger scores correspond to better
partitions, and algorithms are created to find the partition with
the largest score. By far, the most common and used of such quality
functions is modularity~\cite{New06}, that works by comparing the
number of links inside each community to the number of links that
would be expected if the nodes were connected at random, without any
preference for links within or outside the community. A partition
with a large modularity indicates that the communities have many internal
links and few external ones, when compared to a randomized version
of the network. However, despite its success, modularity also has
some shortcomings, decreasing its general usefulness. In this Article,
we study a new quality function, \emph{modularity density}, that was
originally introduced in~\cite{Che13,Che14} and that has been shown
to address the limitations of traditional modularity. We present a
detailed analysis of its properties on synthetic networks typically
used to evaluate quality functions, as well as on random graphs, which
are a commonly used benchmark to test community detection methods.
We also present some limitations that need to be taken into
consideration when using methods based on modularity density.
In addition, we describe a new community detection algorithm based
on this metric, whose computational complexity is quadratic in the
number of nodes, and validate it on synthetic and real-world networks,
showing that it performs better that other currently available methods.
Also, we argue that the nature of modularity density allows for a
direct quantitative comparison of community structures across networks
of different sizes.

\section{Traditional modularity and its limitations}\label{sectrad}
The modularity $Q$ of a network with $N$ nodes and $m$ links is defined as:
\begin{equation*}
Q = \frac{1}{2m}\sum_{ij} \left(A_{ij}-\frac{k_i k_j}{2m}\right)\delta_{C_iC_j}\:,
\end{equation*}
where $A$ is the adjacency matrix of the network,
$k_i$ is the degree of node $i$, $C_i$ is the community
to which node $i$ is assigned and $\delta_{ij}$
is the Kronecker delta. The first term accounts for
the presence or absence of a link between node $i$
and node $j$; the second term, instead, is the expected
number of links between node $i$ and node $j$ in
a random network with the same degree sequence as
the original one.

A first limitation of modularity is that it is intrinsically
dependent on the number and distribution of edges, rather than
on the number of nodes. To see this, denote by $m_C$ and $e_C$
the number of internal and external links of community $C$,
respectively. Moreover, let $k_C=2m_C+e_C$ be the sum of the
degrees of the nodes in community $C$. With this notation, it
is
\begin{equation}\label{modrewr}
Q=\sum_{C\in\mathcal C}\left[\frac{m_C}{m}-\left(\frac{k_{C}}{2m}\right)^2\right]\:,
\end{equation}
where $\mathcal C=\left\lbrace C_1,C_2,\ldots, C_P\right\rbrace$
denotes the set of all communities in the partition. In this expression,
each term in the sum refers to a different community. The first factor of
each term corresponds to the internal density of links in the community,
whereas the second factor encodes the expected density of links in the random
network null model. Now, introduce the positive parameter $\alpha_C$, representing
the ratio of external links to internal ones:
\begin{equation*}
 e_C=\alpha_C m_C\:.
\end{equation*}
The value of $\alpha_C$ is smaller for strong communitites,
and higher for weaker ones. Then, we can write
\begin{equation}\label{eq:modularity}
Q=\sum_{C\in\mathcal C}\left[\frac{m_C}{m}-\left(\frac{2+\alpha_C}{2m}\right)^2 m_C^2\right]\:.
\end{equation}
From this expression, it is clear that a community
$C$ gives a positive contribution to $Q$ only if:
\begin{equation*}
m_C<\frac{4m}{(\alpha_C+2)^2}\:.
\end{equation*}
This implies that the condition for a community
to give a positive contribution only depends
on the number of edges in the community and on
the total number of edges in the network, but not
explicitly on the number of nodes.

A similar result can be obtained considering a network
of $\kappa$ communities disconnected from each other,
along the lines of~\cite{Dan05}. Under the assumption
that all groups have the same number of links, we can
write
\begin{eqnarray*}
m_C &= \frac{m}{\kappa}\:,\\
e_C &= 0\:,\\
k_{C}&= \frac{2m}{\kappa}\:.
\end{eqnarray*}
Then, from~\eref{modrewr}, it is
\begin{equation}\label{eq:mod_k_groups}
Q = \kappa\left[ \frac{1}{m}\frac{m}{\kappa}-\left(\frac{1}{2m}\frac{2m}{\kappa}\right)^2\right]=1-\frac{1}{\kappa}\:.
\end{equation}
This shows that modularity converges to~1 with the number
of communities $\kappa$ regardless of the internal properties
of the communities, such as their size, or the number of
internal edges. As long as $\kappa$ is very large and all
communities have the same number of edges ${}^m/{}_\kappa$,
a network of disconnected trees has the same modularity of
a network of disconnected cliques. As before, we also see
that the number of nodes in each group does not explicitly
contribute to $Q$, and, as an immediate consequence, a network
composed of few cliques has a smaller modularity than a network
composed of many disjoint trees. 

In addition to these results, the effectiveness of modularity
is not constant for all edge densities. To determine its dependence
on this quantity, we follow~\cite{For06} and connect the $\kappa$
groups in a ring configuration, where each community is linked
with exactly one edge to the next one, and one edge to the previous
one in the ring, for a total of $\kappa$ inter-community edges.
In this scenario, we have
\begin{eqnarray*}
m_C &= \frac{m}{\kappa}-1\:,\\
e_C &= 2\:,\\
k_C &= \frac{2m}{\kappa}\:.
\end{eqnarray*}
From~\eref{modrewr}, it follows that
\begin{equation*}
Q = \kappa\left[\frac{1}{\kappa}-\frac{1}{m}-\left(\frac{1}{2m}\frac{2m}{\kappa}\right)^2\right] = 1-\frac{\kappa}{m}-\frac{1}{\kappa}\:.
\end{equation*}
For constant $m$, this expression reaches its maximum when $\kappa=\sqrt{m}$,
for which it is
\begin{equation*}
Q = 1-\frac{2}{\sqrt{m}}\:.
\end{equation*}
Thus, the highest modularity corresponds to a partition
in~$\sqrt{m}$ modules. Once again, the number of nodes
in the communities does not affect its largest possible
value. This major limitation of modularity is known as
the \emph{resolution limit}, and it indicates that modularity,
as a quality function for community detection, has an intrinsic
scale proportional to~$\sqrt{m}$. The number and size of
the communities that can be detected via modularity maximisation
are bound to adhere to this limit, posing a serious question
on the significance of results obtained with this method.
In fact, in a more general framework, Fortunato and Barthélemy~\cite{For06}
have shown that, under some circumstances, the resolution
limit can even force pairs of well-defined communities
to be merged into a larger cluster, because this corresponds
to a higher modularity.

Finally, it is worth noting that the trivial partition
where all the nodes are put together in one single community,
namely the whole network itself, has a modularity of 0.
This can be easily seen from~\eref{eq:modularity},
since in this case the sum has only one term, $\alpha_C=0$
and $m_C=m$, so
\begin{equation*}
Q=\frac{m}{m}-\frac{4m^2}{4m^2}=0\:.
\end{equation*}
At first, this might seem a desirable property for a quality
function, since, intuitively, the trivial partition should not
have a positive modularity. However, this implies that any partition
that achieves a modularity larger than~0 is retained as a valid
community structure. Since community detection algorithms try
to maximize modularity, it is often the case that such a positive
value can be found even on Erdős-Rényi random graphs~\cite{Tre15}.
To stress this point, the trivial partition with $Q=0$ can always
be considered, but since one is interested in the maximum value
of $Q$, it is often discarded in favour of a clustering that
achieves any positive value of modularity. This poses a serious
limitation to the ability of modularity-based algorithms to partition
random graphs correctly.

Several variants of modularity have been proposed
to address the resolution limit. For instance, multi-resolution
methods, such as the one described in~\cite{Are08},
introduce an additional tunable parameter $\eta>0$
in the expression for $Q$:
\begin{equation*}
Q_\eta =\sum_{C\in\mathcal C}\left[\frac{m_C}{m}-\eta\left(\frac{k_C}{2m}\right)^2\right]\:.
\end{equation*}
Larger values of $\eta$ cause $Q_\eta$ to be larger
for partitions with smaller modules, whereas smaller values
favour larger communities. However, this approach suffers
from similar limitations to those presented by the original
modularity~\cite{Lan11}. In particular, $Q_\eta$ has
two contrasting behaviours: small clusters tend to be merged
together, while large communities tend to be split into
subgroups. Networks in which all the communities are of
comparable size are immune to this problem, and one can
find a value of $\eta$ for which they can all be resolved.
However, the existence of an optimal $\eta$ is not guaranteed
in the general case. In particular, for networks whose
community sizes are heterogeneously distributed, e.g.,
following a power law, it is not possible to find a value
of $\eta$ that avoids both problems. The reason for this
is that the nature of the resolution limit is more general
than the specific definitions of modularity and its multi-resolution
extension. Several quality functions for community detection,
including the one just mentioned, can be derived within
the general framework of a first principle Potts model
with Hamiltonian
\begin{equation*}
H = -\sum_{ij}\left[a_{ij}A_{ij}-b_{ij}\left(1-A_{ij}\right)\right]\delta_{C_iC_j}\:,
\end{equation*}
where $a_{ij}$ and $b_{ij}$ are non-negative weights.
Different choices for the weights result in different
quality functions. However, only those using non-local
weights can be truly free from the resolution limit~\cite{Tra11},
while all others, including modularity, multi-resolution
modularity and functions based on quantities such as
betweenness, shortest paths, triangles and loops, can
never avoid it.

\section{Modularity density}\label{secmodden}
Recently, a new quality function called \emph{modularity density}
has been proposed to overcome the issues outlined above~\cite{Che13,Che14}.
Given a network partition, modularity density is defined as
\begin{equation}\label{moddendef}
\fl
Q_{ds} = \sum_{C\in\mathcal C}\left\lbrace\frac{2m_C^2}{mn_C\left(n_C-1\right)}
-\left[\frac{2m_C+e_C}{2m}\frac{2m_C}{n_C\left(n_C-1\right)}\right]^2
-\sum_{\widetilde{C}≠C}\frac{m_{C\widetilde{C}}^2}{2mn_Cn_{\widetilde{C}}}\right\rbrace\:,
\end{equation}
where $n_C$ is the number of nodes in community
$C$, the internal sum is over all communities different
from $C$, and $m_{C\widetilde{C}}$ is the number of
edges between community $C$ and community $\widetilde{C}$.
This new metric brings two major improvements over
traditional modularity. First, it contains an explicit
penalty for edges connecting nodes in different
communities. This addresses the problem of the
splitting of large communities, since each split
introduces external links and is thus penalized.
Second, all terms, including the penalty for inter-community
edges, are explicitly weighted by the community
sizes. Therefore, a partition with many edges linking
two small communities is penalized more than one
with the same number of edges linking two large
ones. Thus, modularity density introduces local
dependencies that are not found in traditional
of modularity. Additionally, it is not related
to the Potts model Hamiltonian, thus
avoiding the resolution limit problem. Note that~\eref{moddendef}
requires $n_C>1$, which implies that partitions
with communities consisting of an isolated node
are not allowed.

To investigate the properties of modularity density
in more depth, rewrite the expression for $Q_{ds}$ as
\begin{equation}\label{eq:mod_dens}
Q_{ds} = \sum_{C\in\mathcal C}\left[\frac{m_C}{m}p_C-\left(\frac{2m_C+e_C}{2m}p_C\right)^2
-\sum_{\widetilde{C}≠C}\frac{m_{C\widetilde{C}}^2}{2m n_C n_{\widetilde{C}} }\right]\:,
\end{equation}
where
\begin{equation*}
p_C = \frac{2m_C}{n_C\left(n_C-1\right)}\:.
\end{equation*}
The parameter $p_C$ can assume values between~0
and~1, since it is the fraction of possible internal
links actually present in community~$C$. Thus,
it measures the connection density of the community,
or, equivalently, the probability that two random
nodes inside $C$ are connected. From~\eref{eq:mod_dens}
it is clear that having many internal edges is
not enough for a community to give a large contribution
to modularity density. In fact, a strong community
is one where the density of edges, rather than
their number, is large. This also agrees with the
intuitive notion that a community is a group of
nodes that are densely connected amongst each other.
Thus, a good partition is one that is characterized
at the same time by a large number of intra-community
links and a high density of edges within the communities.
Modularity density achieves this by accounting
for the number of nodes in each group and, in this
sense, it has a more natural dependence on the
local properties of the network and of the partition
under consideration than does traditional modularity.

Next, it is instructive to study the behaviour of modularity
density in the same cases described in the previous section.
First, consider a network partitioned in just two communities,
$C$ and $\widetilde{C}$. The contribution to $Q_{ds}$ of community
$C$ is:
\begin{equation*}
Q_{ds}^C = \frac{m_C}{m}p_C-\left(\frac{2m_C+e_C}{2m}p_C\right)^2-\frac{m_{C\widetilde{C}}^2}{2mn_Cn_{\widetilde{C}}}\:.
\end{equation*}
Introducing the proportionality constant $\alpha_C$ as before, it is
\begin{equation*}
Q_{ds}^C = \frac{m_C}{m}\left[p_C-\frac{\left(2+\alpha_C\right)^2}{4m}p_C^2 m_C-\frac{\alpha_C^2 m_C}{2n_C\left(N-n_C\right)}\right]\:,
\end{equation*}
where we used $n_{\widetilde{C}}=N-n_C$ and $m_{C\widetilde{C}}=e_C$.
Unlike what happens with traditional modularity, the contribution of
a single community depends explicitly on the number of internal links
\emph{and} on the size of the community itself.

\begin{figure}[t]
\centering
\includegraphics[width=0.75\textwidth]{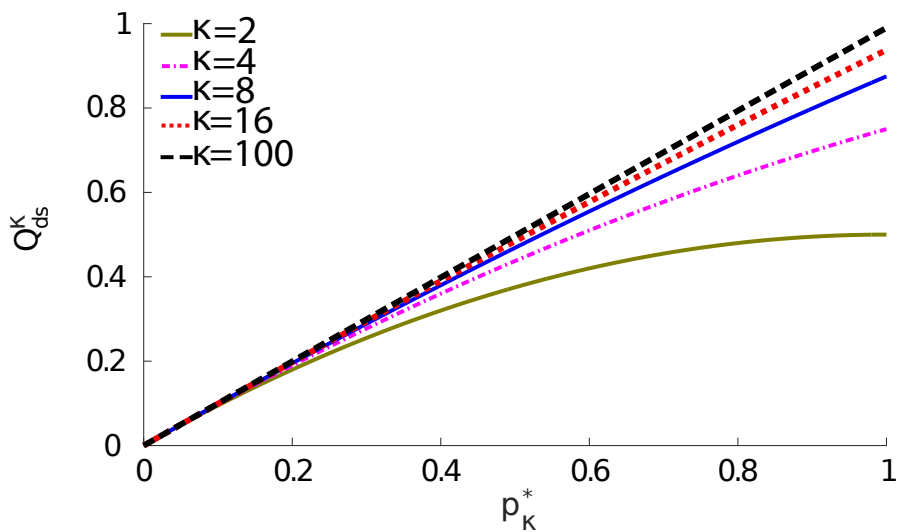}
\caption{Modularity density increases with number
of communities and their edge density. In networks
composed of $\kappa$ disconnected modules, the modularity
density $Q_{\mathrm{ds}}^{\kappa}$ depends only on
$\kappa$ and the edge density of the communities
$p_{\kappa}^{\star}$. Fixing one of the two parameters,
$Q_{\mathrm{ds}}^{\kappa}$ always increases with
the other.}\label{fig1}
\end{figure}
Consider now again a network composed of $\kappa$ disjoint
communities. Assuming that each community has the same number
of nodes ${}^N/{}_\kappa$ and the same number of edges ${}^m/{}_\kappa$,
the modularity density of such a network is:
\begin{equation}\label{eq:mod_dens_k_groups}
Q_{ds}^\kappa = \kappa\left[\frac{p_{\kappa}^{\star}}{\kappa}-\left(\frac{p_{\kappa}^{\star}}{\kappa }\right)^2\right]=p_\kappa^\star\left(1-\frac{p_\kappa^\star}{\kappa}\right)\:,
\end{equation}
where
\begin{equation*}
p_\kappa^\star=\frac{2m}{N\left(\frac{N}{\kappa}-1\right)}
\end{equation*}
is the connection density of the communities.
The first major difference between~\eref{eq:mod_k_groups}
and~\eref{eq:mod_dens_k_groups} is that $Q_{ds}^\kappa$
depends not only on the number of communities,
but also on their density of edges, unlike traditional
modularity, which only depends on $\kappa$. Also,
for a fixed value of $\kappa$, $Q_{ds}^\kappa$
increases with $p_\kappa^\star$ (see~\fref{fig1}).
This is remarkable, since it indicates that the
strength of the partition increases as more
links are added within each group, in striking
opposition with the behaviour of traditional
modularity. We also note that for a fixed value
of $p_\kappa^\star$, modularity density increases
with the number of communities. Its theoretical maximum is
reached in the limiting case of an infinite number
of communities, with the special requirement
that they are all cliques. Moreover, in one more
substantial difference with traditional modularity,
a network composed of few cliques in general has
a higher modularity density than a network
composed of an infinite number of sparse communities.

Finally, we study the test case of the ring of $\kappa$
communities each linked by a single edge to the next community
and a single edge to the previous one. As before, it is
$m_C={}^m/{}_\kappa-1$ and $e_C=2$, In addition, $m_{C\widetilde{C}}=1$
and $n_C={}^N/{}_\kappa$. Introducing the variables
\begin{equation*}
\beta_\kappa = \frac{\frac{m}{\kappa}-1}{m}
\end{equation*}
and
\begin{equation*}
p_\kappa^\star = \frac{2(\frac{m}{\kappa}-1)}{\frac{N}{\kappa} (\frac{N}{\kappa}-1)}\:,
\end{equation*}
we can write the modularity density as
\begin{equation}\label{eq:mod_dens_ring_network}
Q_{ds}^{\mathrm{ring}}=\kappa\left[\beta_\kappa p_\kappa^\star-\left(\beta_\kappa+\frac{1}{m}\right)^2 \left(p_\kappa^\star\right)^2 -\frac{\kappa^2}{m N^2}\right]\:.
\end{equation}
The optimal number of communities is the one that maximizes
this expression, or, equivalently, the one for which its derivative
vanishes. Differentiating $Q_{ds}^{\mathrm{ring}}$ with respect
to $\kappa$, we obtain
\begin{equation*}
\fl\frac{\partial Q_{ds}^{\mathrm{ring}}}{\partial\kappa} = \kappa\beta_\kappa\partial_\kappa p_\kappa^\star-2\left(\beta_\kappa +\frac{1}{m}\right)p_\kappa^\star\partial_\kappa p_\kappa
+\left(\beta_\kappa +\frac{1}{m}\right)^2 (p_\kappa^\star)^2 + \left(\beta_\kappa-\frac{1}{\kappa}\right)p_\kappa -\frac{3 \kappa^2}{m N^2}\:,
\end{equation*}
with
\begin{equation*}
\partial_\kappa p_\kappa^\star = \frac{2 \left(\kappa^2 -2 \kappa N+m N\right)}{N (\kappa-N)^2}\:.
\end{equation*}
This expression does not have a simple general
root in terms of $\kappa$. Rather, the solutions
depend on the local and global properties of the
network. Thus, the number of groups does not seem to be
constrained by an intrinsic scale of order $\sqrt{m}$.

As briefly discussed above, a major drawback of traditional
modularity is that algorithms based on its maximization often
find supposedly viable partitions on graphs with no ground-truth
community structure. In such cases, the correct partition is
either the one where all nodes are placed together, or the
one with $N$ communities, each consisting of a single node.
In either case, modularity vanishes. Thus, modularity-maximizing
algorithms often suggest spurious community structures simply
beause they have a non-zero modularity. Conversely, from~\eref{eq:mod_dens}
it follows that the one-group partition has a modularity density
\begin{equation*}
 Q_{ds}^1=p\left(1-p\right)\:,
\end{equation*}
where $p$ is the network density. Note that this expression
is a parabola, whose roots are $p=0$ and $p=1$, which are the
fully disconnected and fully connected graphs, respectively.
Thus, a partition's $Q_{ds}$ needs not only to be positive,
but also to lie above the parabola for an algorithm based on
modularity density maximization to accept it. We will see that
this makes such algorithms not find communities on random graphs,
as should be the case for a reliable community detection method.

\section{A modularity density maximisation algorithm}
Having discussed the advantages of modularity density
as a quality function, we propose a community detection
algorithm based on its maximization. Currently, the only
published modularity density algorithm~\cite{Che14} is
based on iterations of two steps, namely splitting and
merging. The algorithm is divisive, starting from a partition
where all the nodes are placed in a single community
and then using bisections. Each splitting is performed
using the Fiedler vector of the network, which is the
eigenvector of the graph Laplacian corresponding to the
second smallest eigenvalue. The graph Laplacian $L$ is
defined as $L=D-A$, where $D$ is the diagonal matrix
of the node degrees. The merging steps try to merge pairs
of communities together if doing so improves the current
partition. The two steps are repeated until the partition
cannot be improved any longer, and the algorithm is deterministic,
meaning that the same initial network always yields the
same partition. Here, we extend and adapt an existing
modularity maximisation algorithm, originally proposed
in~\cite{Tre15}, which achieves the largest published
scores of traditional modularity. Along the lines of
the original method, our algorithm consists of four main
steps, which we describe below. \ref{imple}
contains a fully detailed discussion of the algorithm
implementation and its computational complexity.

\subsection{Bisection} In this step, we try to bisect
the community under consideration (see~\fref{fig2}A). To do so, we use
the leading eigenvector of the modularity matrix. Despite
suffering from the limitations discussed above, modularity
still provides a good initial guess for a partition
that is then refined by the subsequent steps.

\subsection{Fine Tuning}
\begin{figure}[t]
\centering
\includegraphics[width=0.75\textwidth]{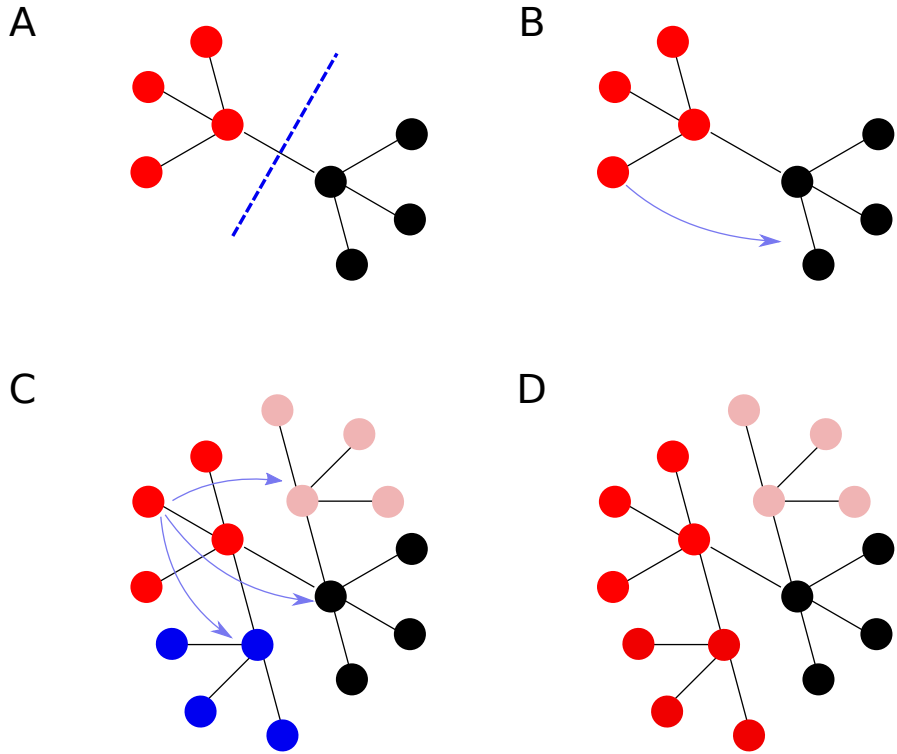}
\caption{Schematic illustration of the algorithm steps.
A) Bisection considers splitting a group of nodes into two
separate communities.
B) Fine tuning considers moving each node of a newly
bisected group from its current community to the other.
For example, it considers moving the red node to the black
community.
C) Final tuning considers moving each node of the network
from its current community to any other existing community.
For example, it considers moving the red node to the blue,
the black and the pink communities.
D) Agglomeration considers grouping two separate communities
into a single one. For example, it considers grouping the red
and blue nodes of panel~C into a single red community.}\label{fig2}
\end{figure}
After every bisection, the partition
can be often improved by using a variant of the Kernighan-Lin
algorithm~\cite{Ker70}. We consider moving every node $i$ from
the community into which it was assigned to the other (see~\fref{fig2}B). Every
such move would result in a change $\Delta Q_{ds}^{i}$ of the
quality function, and we perform the move yielding the largest
of such changes $\Delta_{\max}Q_{ds}^i$. Note that we introduce
here a non-deterministic factor: given a tolerance parameter
$\tau_{acc}$, we consider all moves achieving a change of modularity
density within the interval $\left[\Delta_{\max}Q_{ds}^i-\tau_{acc},\Delta_{\max}Q_{ds}^i\right]$
to be equivalent. If more than one move falls within the acceptance
interval, we randomly choose one to accept. This stochasticity
allows the algorithm to explore the partition space without
getting stuck on a local maximum, since it can accept moves
that are not always optimal. Once a move has been performed,
the corresponding node is flagged as blocked. Then, every non-blocked
node is considered again and the procedure is repeated, until
all nodes have been considered. At the completion of an iteration
of this step, a decision tree is formed where each node of the
tree represents a sequence of nodes in the network switching
community, with an associated $\Delta Q_{ds}$ equal to the sum
of all the changes in modularity density along the branches
leading to the tree node. Then, we randomly choose a node in
the decision tree amongst those achieving the largest positive
increase in modularity density within an interval determined
by the tolerance parameter $\tau_{acc}$, and perform all the
moves corresponding to the chosen node. Finally, the whole step
is repeated until no improvement in $Q_{ds}$ can be obtained.

\subsection{Final Tuning} A further refinement of a current
partition can be achieved by performing an additional tuning
step. In the final tuning, we consider every node $i$ and
try to move it to every other possible community $C$ already
present in the partition (see~\fref{fig2}C). The step is performed in a similar
fashion to the fine tuning, repeatedly considering all the moves which
result in an increase of modularity density in a small interval
defined by the tolerance parameter $\tau_{acc}$ until all nodes
have been moved. As before, we build a decision tree of partial
switches and then perform all the moves up to the level in the
tree that has been selected amongst those yielding the largest
increase in $Q_{ds}$. We repeat this step until no further refinements
can be found.

\subsection{Agglomeration} A step that merges pairs of communities
is fundamental. First of all, unlike both tuning steps, which are
local because they only consider moving one node at a time, merging
communities is a non-local step that allows one to better explore
the landscape of modularity density~\cite{Tre15}. For example, merging
two entire communities can result in an increase of the quality function
while partial mergers, i.e., moving only some nodes from one community
to the other, could still have a lower score than the starting partition.
Therefore, using only local moves, one could discard those partial mergers
because they temporarily decrease the partition score, thus
never achieving the beneficial complete merging of the two communities.
In the case of modularity density, the agglomeration step is even
more important, since no series of local moves could ever produce
the full merging of two communities. This happens because modularity
density does not allow communities of size~1. Thus, even if local
steps had succeeded in moving all nodes except two from one community
to another, any further move would be prohibited because it would
result in a single-node community. This makes a global move essential
for our algorithm. In the agglomeration step, we consider pairs of
communities $C$ and $\widetilde{C}$ and try to merge them (see~\fref{fig2}D). Each move
results in a change in modularity density $\Delta Q_{ds}^{C,\widetilde{C}}$
and we randomly choose the move amongst those in the interval
$\left[\Delta_{\max}Q_{ds}^{C,\widetilde{C}}-\tau_{acc},\Delta_{\max}Q_{ds}^{C,\widetilde{C}}\right]$,
where $\Delta_{\max}Q_{ds}^{C,\widetilde{C}}$ is the largest increase
in modularity density achieved by any move. We build a decision tree
by progressively merging pairs of communities, until there is only
a single community left. We then look at the nodes in the tree corresponding
to the largest increase in modularity density but, in difference from
the previous steps, if more than one node results in the same increase,
we select the one with the smallest number of communities. The whole
step is repeated until the current partition cannot be improved further.

\subsection{Summary}
\begin{figure}[t]
\centering
\includegraphics[width=0.75\textwidth]{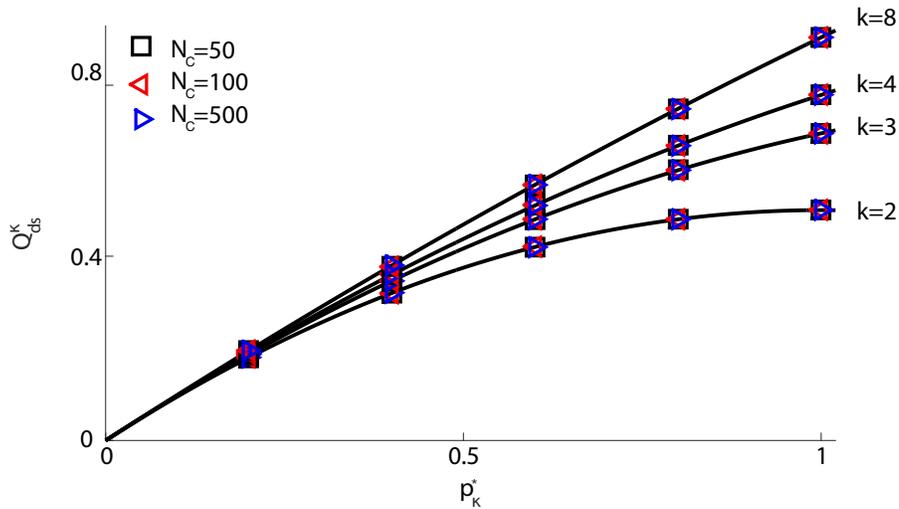}
\caption{Modularity density for networks of $\kappa$ disconnected communities.
The predictions of~\eref{eq:mod_dens_k_groups} (solid lines) are confirmed by
numerical simulations throughout the range of $p_\kappa^\star$ and for different
values of $\kappa$. For each $\kappa$, we consider groups with 50, 100 and 500
nodes, respectively. Additionally, and as expected, we observe that the value
of modularity density does not depend on the number of nodes in each community,
but only on the number of communities and their internal density. Each point is
the average over 100 network realizations.}\label{fig3}
\end{figure}
With these four steps, the algorithm can be summarized as:
\begin{itemize}
\item Start with a single community containing all nodes.
\item Try to bisect the network using the leading eigenvector of the modularity matrix.
\item If the bisection was successful, then perform a fine tuning step.
\item Iterate the bisection and fine tuning steps on each of the communities in the current partition, until no further splitting and refinement can be performed.
\item Perform the final tuning step.
\item Perform the agglomeration step.
\item Repeat the sequence of steps until it is no longer possible to find an increase in modularity density.
\end{itemize}
As described in detail in~\ref{imple}, the worst-case
computational complexity of the full algorithm is $O\left(N^2\right)$.

\section{Validation}
To validate our algorithm, we test it on several
synthetic and real-world networks. First, we verify
that it reproduces the theoretical predictions on
networks of disconnected communities and on rings
of modules, discussed in~\sref{sectrad} and~\sref{secmodden}.
Then, we analyze its behaviour on random networks
belonging to different ensembles. Finally, we run
it on a set of benchmark networks, comparing the
results with the best ones currently published.

\subsection{Disconnected communities and rings}
\begin{figure}[t]
\centering
\includegraphics[width=0.45\textwidth]{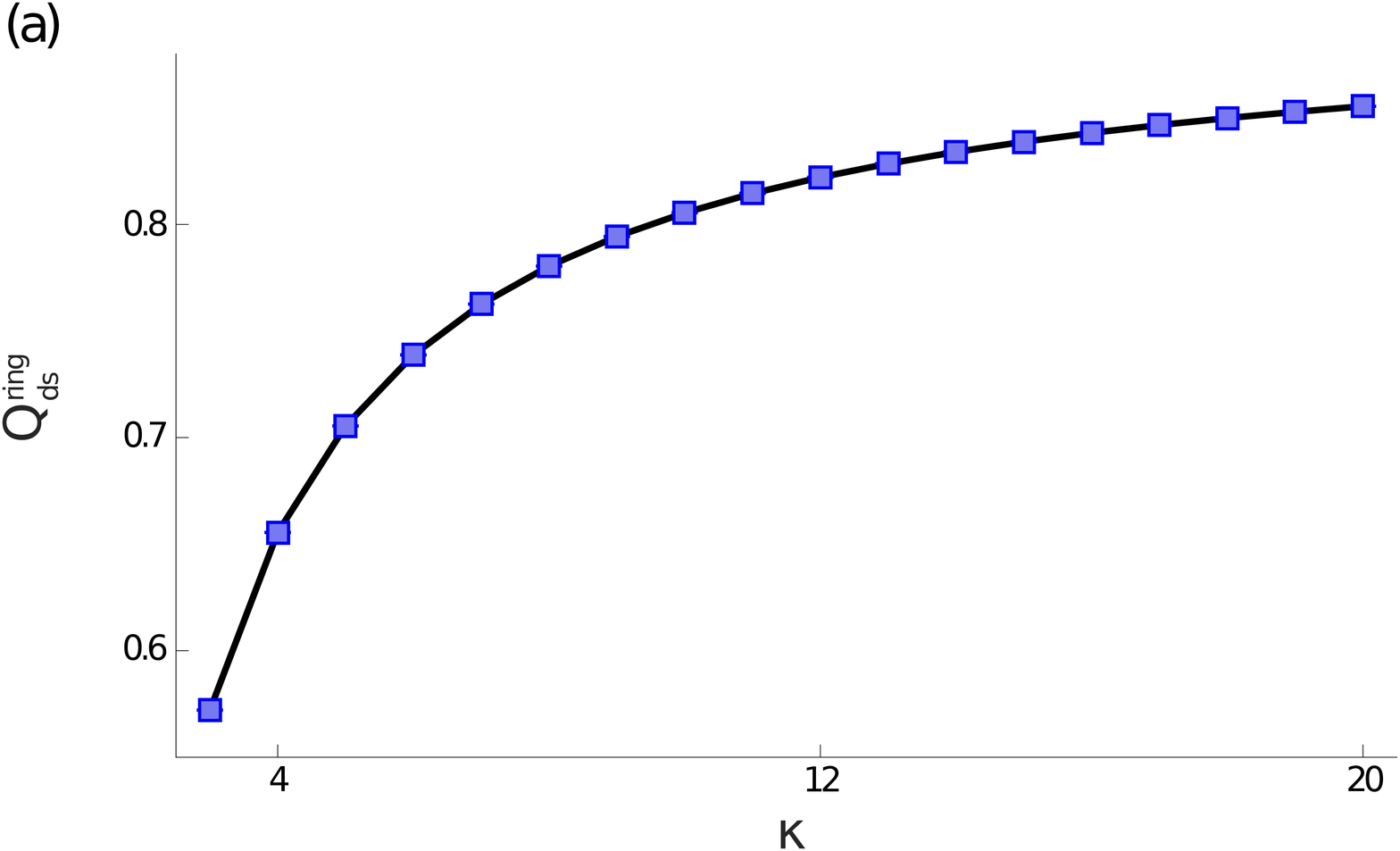}
\includegraphics[width=0.45\textwidth]{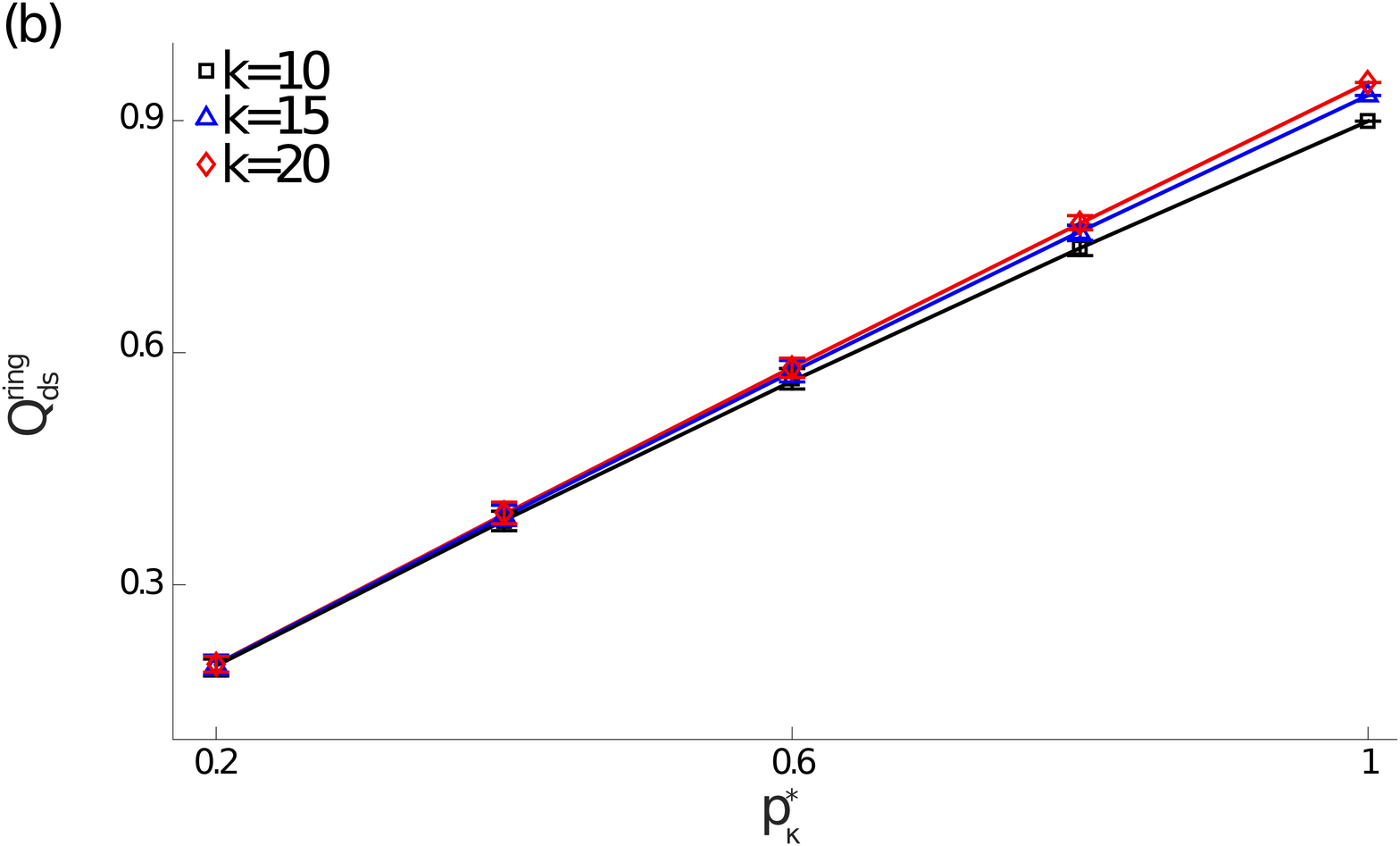}
\caption{Modularity density for rings of communities.
We simulate ring networks composed by with a varying
number of communities $\kappa$ and compare the theoretical
values of modularity density with the results of our
algorithm. In panel~(a) we consider networks of $\kappa$
fully connected cliques, finding a perfect agreement
between theoretical value (solid line) and simulations
(squares). In panel~(b), we build networks with different
fixed values of $\kappa$ and vary their internal density.
Note that, differently from~(a), here the groups are
not fully connected. The theoretical values (lines) and
simulation results match precisely. In both panels, each
point is the average over 100 realizations of the same
network.}\label{fig4}
\end{figure}
First, we consider networks formed by $\kappa$
disconnected communities. \Eref{eq:mod_dens_k_groups}
indicates that the modularity density of such
networks depends only on the connection probability
$p^{\star}_{\kappa}$ and on $\kappa$ itself, but
not on the size of each community. We find an
exact agreement between the simulation results
and the theoretical prediction for all the values
of $\kappa$ (\fref{fig3}). We also note that the
values of modularity density found in the simulations
do not depend on the number of nodes in the communities.

As a second test, we simulate two types of ring networks
of communities. We start by making the communities cliques
of 5 fully connected nodes, and vary $\kappa$ from 3 to
20. From~\eref{eq:mod_dens_ring_network}, the expected modularity
density of these networks is
\begin{equation*}
Q_{ds}^{\mathrm{ring}} = \kappa\left[\beta_\kappa-\left(\beta_\kappa+\frac{1}{m}\right)^2-\frac{\kappa^2}{25m}\right]\:.
\end{equation*}
The comparison between the modularity density predicted
by this expression and the values obtained in our simulations
is shown in~\fref{fig4}(a). We find a precise agreement between
the two, showing that our algorithm correctly identifies
the cliques without splitting them, and finds the right
value of modularity density. Next, we build ring networks
in which we fix $\kappa$ and vary the community density
$p_\kappa^\star$. Each community contains 50 nodes, and
we vary $p_\kappa^\star$ from $0.2$ to 1, performin the
test for $\kappa=10$, $\kappa=15$ and $\kappa=20$. The
results, in~\fref{fig4}(b), show a perfect agreement in all
cases, again indicating that our algorithm correctly partitions
the networks.

\subsection{Random networks}
\begin{figure}[t]
\centering
\includegraphics[width=0.75\textwidth]{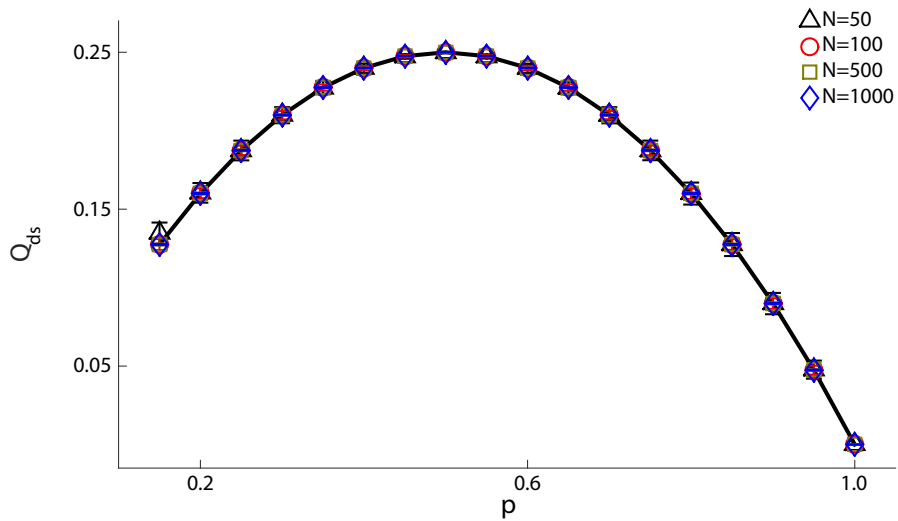}
\caption{Modularity density on Erdős-Rényi graphs.
We build ensembles of random networks, with different
sizes and different densities, comparing the theoretical
modularity density (solid line) and the one found
by our algorithm. Up to finite-size effects for the
smallest and least dense networks, we find a perfect
agreement between theoretical prediction and simulation
results, with all the results collapsing on the same
curve. Each point is the average over 1000~realizations
of the same network parameters.}\label{fig5}
\end{figure}
As we argued in the previous sections, a desirable
feature of a community detection algorithm is that
it does not propose a complex partition of graphs
without ground-truth community structure. To verify
that our algorithm satisfies this requirement, we
test it on Erdős-Rényi random graphs. For graphs in
this ensemble, every possible edge between $N$ nodes
exists independently with probability $p$. Thus, the
average number of edges is $\frac{1}{2}Np\left(N-1\right)$.
These networks do not have any true community structure,
since all their edges are fully random, and
thus they are one of the benchmarks against which
community detection algorithms are often tested.
For our simulations, we create networks with values
of $p$ from $0.15$ to $0.90$ and number of nodes 50,
100, 500 and 1000. The results, in~\fref{fig5}, show
that for all network sizes, the average modularity
density matches almost perfectly the theoretical prediction.
Even for small networks, where finite-size effects
are largest, the values lie in close proximity to
the theoretical parabola and we can only observe a
small deviation for the smallest networks at low values
of $p$. Also note that all the results collapse on
the theoretical curve, which does not depend on network
size. These results represent a major improvement
over modularity-based algorithms, that typically detect
communities even on Erdős-Rényi networks. In
addition, Erdős-Rényi networks are locally tree-like for
low enough values of $p$, and highly clustered for $p$ close
to~1. Thus, the results also indicate that modularity density
is highly effective in detecting when no real communities
exist in locally tree-like graphs, and does not introduce
spurious modules even when the clustering increases. In
fact, also the limiting case of fully connected graphs,
which corresponds to a link probability $p$ identically
equal to~1, is properly identified by our method.

\begin{figure}[t]
\centering
\includegraphics[width=0.75\textwidth]{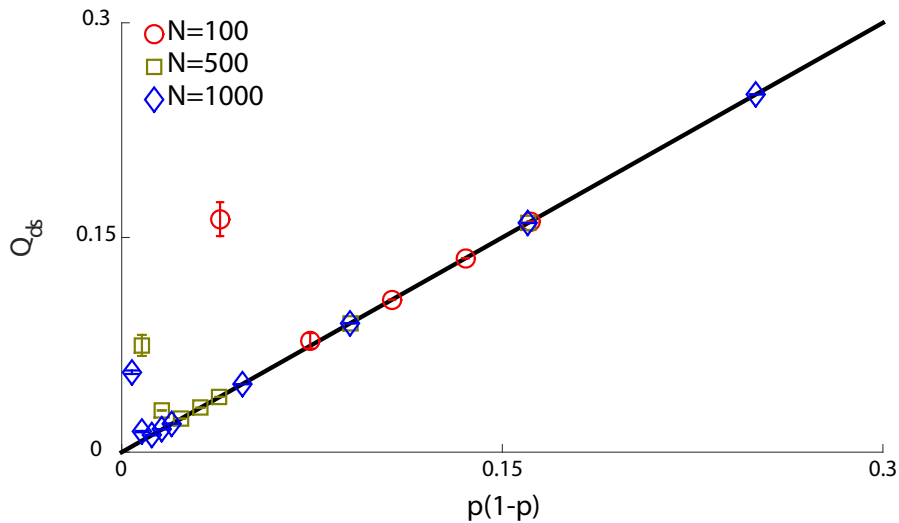}
\caption{Modularity density on random regular graphs.
We build ensembles of random regular networks, with different
sizes and degree, comparing the theoretical modularity
density (solid line) and the one found by our algorithm.
With the exception of the sparsest network for each size,
we find a good agreement between theoretical prediction
and simulation results, with all the results collapsing
on the same curve. Each point is the average over 100~realizations
of the same network parameters.}\label{fig6}
\end{figure}

As a special case
of random networks, we also study random regular graphs.
Random regular graphs are random networks where
all nodes have the same degree, but the edges are still
randomly placed. Using the algorithm described in Ref.~\cite{del10},
we create random regular graphs with~100, 500 and~1000
nodes. For each of the three network sizes, we consider
degrees ranging from~4 to~20, 100 and~500, respectively.
For every pair of size and degree, we generate 100~network
realizations, on which we run our community detection algorithm.
The results, depicted in~\fref{fig6}, show a good agreement
between theoretical predictions and simulations. The only exceptions
are three cases that correspond to the sparsest graph of each given size.

\begin{figure}[t]
\centering
\includegraphics[width=0.75\textwidth]{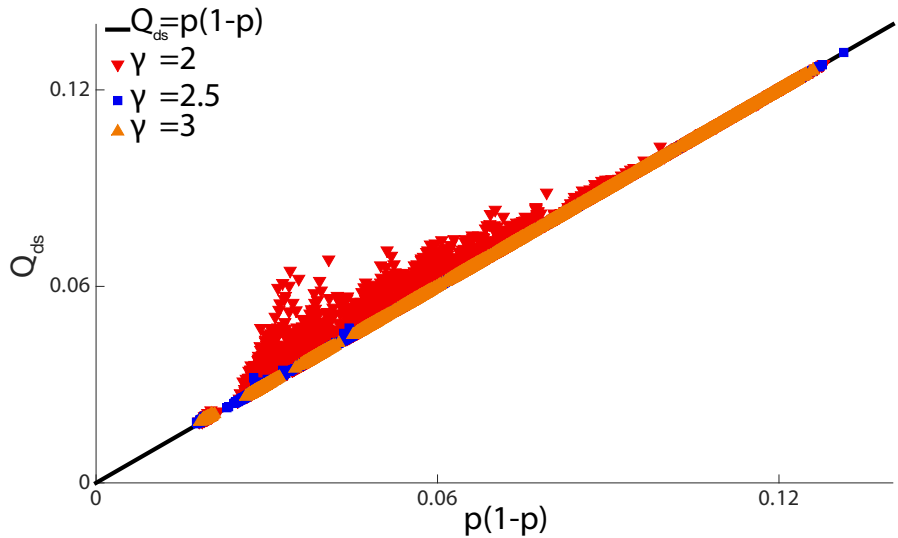}
\caption{Modularity density for LFR networks.
We run our algorithm on random LFR networks without
community structure, with $N=500$ nodes and varying
parameters. In particular, we let the mean degree
$\left\langle k\right\rangle$ assume the values
15, 25, 35, 44 and 55, and the largest degree
$k_{\max}$ be 150, 200 and 250. For each combination
of the parameters, we generate 100 networks
and for each we record the edge density $p$ and
the largest modularity density our algorithm
finds. The plot shows considerable agreement
between the theoretical modularity density (solid
line) and the one found by the algorithm. The
only deviations appear for $\gamma=2$ and low $p$,
and they are probably due to the breakdown of the
LFR model
for this limiting value of the
degree distribution exponent.}\label{fig7}
\end{figure}

These results show one of the major strengths
of modularity density. However, it is well known that most real-world networks are not well
represented by Erdős-Rényi graphs or random regular graphs. Rather, they
are characterized by heterogeneous degree distributions.
Thus, to further verify the performance of our algorithm,
we test it on LFR networks~\cite{Lan08}. These constitute
a set of widely-used benchmark networks, whose distributions
of degrees and community sizes follow a power-law $P\left(k\right)\sim k^{-\gamma}$.
For our tests, we fix the network size to $N=500$ and
vary the other parameters, namely the exponent $\gamma$
of the degree distribution, the mean degree $\left\langle k\right\rangle$
and the largest degree $k_{\max}$. Also, we ensure that
the networks thus created contain a single community,
so that no actual community structure is present. We
run our algorithm on the networks thus generated and
compare its results with the theoretical expectations.
The results, presented in~\fref{fig7}, show that for $\gamma=2.5$
and $\gamma=3$, the modularity density found by the algorithm
closely follows the predicted value for networks of all
densities. We do observe, however, some deviations
from the predicted values at $\gamma=2$. This is probably
due to the fact that, asymptotically, no networks exist
with a pure power-law degree distribution for $\gamma<2$~\cite{del11}.
Thus, in the limit of $\gamma=2$, and particularly for
low densities, a spurious structure of stars with bridges
appears, effectively introducing communities in the networks.

\subsection{Benchmark networks}
We now verify the performance of our algorithm
on some well known networks, for which results
of the maximum modularity density obtained so
far are available. The first is Zachary's Karate
Club network~\cite{Zac77}. This is a friendship
network between 34 members of a karate club in
a U.S. university during the 1970s and it has
become one of the most standard benchmarks to
test community detection algorithms. The interest
in this network lies in the fact that, not long
after it was recorded, the club split into two
subgroups due to internal problems between two
members, namely the manager and the coach. Thus,
a traditional challenge is to be able to detect
these two groups based only on the friendship
data available in the network topology, under
the assumption that the members would decide to
follow whichever leader they were more strongly
related to between the coach and the manager.
Of the 561 possible edges in the network, only
78 of them are present, making the network fairly
sparse, with an effective connection probability
$p\approx 0.139$.

A second benchmark network we consider is the American
College Football Club network~\cite{Gir02}. Here, the
nodes represent different college football clubs and an
edge connects two teams if there has been a regular-season
game between them during the 2000 season. This network
is known to have a natural community structure because
the teams are divided into different leagues, thus making
matches between teams more or less likely depending on
the group they belong to.

\begin{table}[t!]
\caption{Accuracy validation. The comparison
between the published results and the ones obtained with
our algorithm on real-world and synthetic benchmark networks
shows that our algorithm always performs better than the
current best one. Note that currently this is the only other
algorithm based on modularity density maximisation.
All the already published results are
found in~\cite{Che14}.}\label{tab1}
\begin{indented}
\item[]\begin{tabular}{@{}lllll}
\br
Benchmark & $Q_{ds}$ & $Q_{ds, pub}$ & $p$ & $p(1-p)$\\
\mr
Karate Club     & $0.235$            & $0.231$  & $0.139$            & $0.120$           \\
Football Club   & $0.490931$         & $0.4909$ & $0.0935$           & $0.0848$          \\
LFR, $\mu=0.05$ & $0.5220\pm 0.0039$ & $0.4979$ & $0.0156\pm 0.0001$ & $0.0154\pm 0.0001$\\
LFR, $\mu=0.10$ & $0.4638\pm 0.0033$ & $0.4522$ & $0.0154\pm 0.0001$ & $0.0152\pm 0.0001$\\
LFR, $\mu=0.15$ & $0.4249\pm 0.0030$ & $0.4013$ & $0.0157\pm 0.0002$ & $0.0155\pm 0.0002$\\
LFR, $\mu=0.20$ & $0.3982\pm 0.0054$ & $0.384$  & $0.0156\pm 0.0001$ & $0.0154\pm 0.0001$\\
LFR, $\mu=0.25$ & $0.3465\pm 0.0085$ & $0.3347$ & $0.0156\pm 0.0001$ & $0.0154\pm 0.0001$\\
LFR, $\mu=0.30$ & $0.2986\pm 0.0034$ & $0.2619$ & $0.0156\pm 0.0001$ & $0.0154\pm 0.0001$\\
LFR, $\mu=0.35$ & $0.2546\pm 0.0101$ & $0.2377$ & $0.0156\pm 0.0001$ & $0.0154\pm 0.0001$\\
LFR, $\mu=0.40$ & $0.2340\pm 0.0069$ & $0.199$  & $0.0156\pm 0.0001$ & $0.0154\pm 0.0001$\\
LFR, $\mu=0.45$ & $0.2029\pm 0.0064$ & $0.169$  & $0.0156\pm 0.0001$ & $0.0154\pm 0.0001$\\
LFR, $\mu=0.50$ & $0.1579\pm 0.0027$ & $0.1385$ & $0.0156\pm 0.0001$ & $0.0154\pm 0.0001$\\
\br
\end{tabular}
\end{indented}
\end{table}
Finally, we consider again some LFR benchmark networks,
choosing a set of parameters for which already published
results exist. \Tref{tab1} presents a comparison between
the results obtained using our algorithm and the best results
available in the literature. Note that currently there is only
one other algorithm based on modularity density. Because of the stochasticity
within our method, for each value of the mixing parameter
$\mu$, we create 10 realizations of the network and run
the algorithm 100 times on each, reporting the average
maximum modularity density found. In all cases considered,
our algorithm finds a partition with higher modularity
density than the best one currently published.

\section{Limitations}
So far, we have shown that our algorithm identifies
the correct value of modularity density on a range
of test networks. However, it is also worth noting
that methods based on modularity density present some
limitations in special cases.

\begin{figure}
\centering
\includegraphics[width=0.75\textwidth]{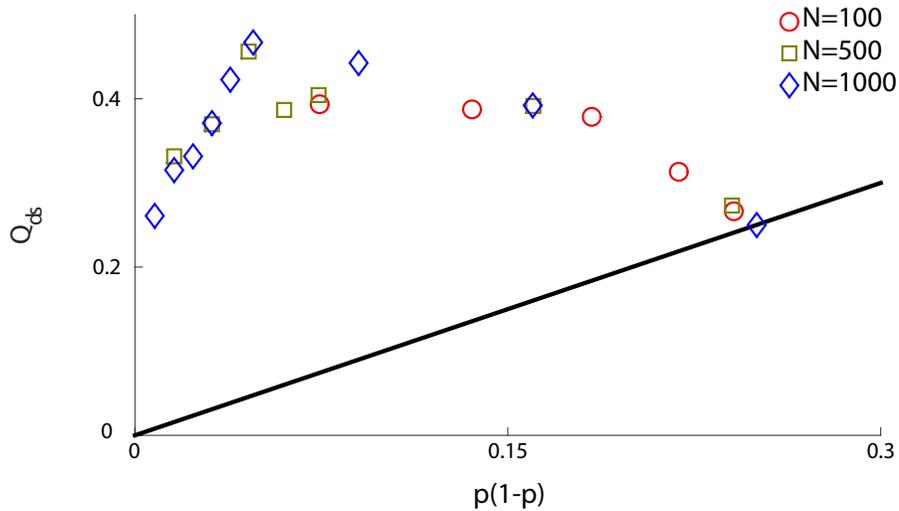}
\caption{Modularity density on regular ring lattice graphs.
We build regular ring lattice graphs, with different
sizes and degrees, comparing the theoretical modularity density
(solid line) and the one found by our algorithm. The results
show that ring lattices are a limitation of modularity density.
In fact, in almost all cases it is possible to find a network
partition with a higher modularity density than the trivial
one, corresponding to the absence of ground-truth communities.}\label{fig8}
\end{figure}

To show this, we analyze regular ring lattice networks.
These are graphs composed by a ring of $N$
nodes, each connected to a number of neighbours in each
direction. We consider networks of~100, 500 and~1000 nodes,
for which the number of neighbours of each node varies
from~8 to~40, 200 and~500, respectively.
The results, depicted in~\fref{fig8},
illustrate how the theoretical prediction and the simulated results differ
on almost all cases considered. In fact, with one exception,
there is always a set of communities with a higher value of modularity
density than the one corresponding to the trivial partition.

\begin{figure}
\centering
\includegraphics[width=0.75\textwidth]{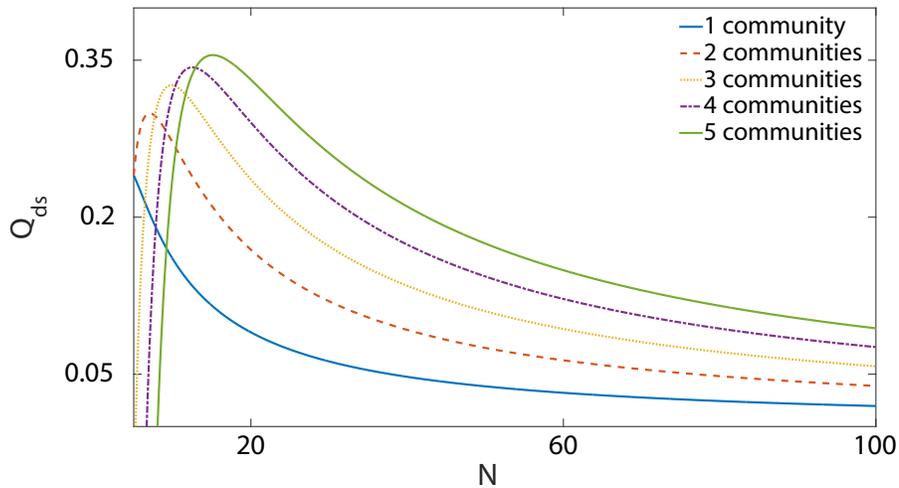}
\caption{Modularity density
for random trees partitioned into equally sized communities.
The values indicate that trees constitute a limitation to the
application of modularity density, since, for more than~6 nodes,
partitions with more than one community always have larger modularity
density than the trivial one.}\label{fig9}
\end{figure}

Trees are another special case where modularity density exhibits
some shortcomings. To see this, consider a
tree with $N$ nodes. If all the nodes are put in a single
community, the modularity density is given by:
\begin{equation*}
Q^{(1)}_{ds}=p(1-p)=\frac{2(N-2)}{N^2}
\end{equation*}
If instead one divides the nodes between two different communities
of equal size and equal number of internal links, it is
\begin{equation*}
Q^{(2)}_{ds}=\frac{4(1+(N-4)N)}{N^2(N-1)}\:,
\end{equation*}
where we used the fact that, for a tree, there can only be one link between the
two communities if they do not consist of disconnected components. It follows that,
for $N\geqslant 6$, $Q^{(2)}_{ds}>Q^{(1)}_{ds}$, that is,
for trees with more than 6~nodes, a partition
in two equally sized communities always has a larger modularity
density than the trivial partition. More generally, partitions
with a larger number of equally sized communities tend to have
a larger score, as can be seen in~\fref{fig9}.

\section{Conclusions}
Communities are a fundamental structure that is often present
in real-world complex networks. Thus, the ability to accurately
and efficiently detect them is of great relevance to the analysis
of complex data sets. Despite their success, traditional methods
based on modularity have been shown to suffer from limitations.
We have presented a detailed analysis of the properties of modularity
density, an alternative quality function for community detection,
showing that it does not suffer from the drawbacks that affect
traditional modularity. In particular, modularity density does
not depend separately on the size of the network or the number
of edges, but only on the combination of these two properties
in terms of the density of links within the communities. As a
consequence, it allows a direct quantitative comparison of the
community structure across networks of different sizes and number
of edges. At the light of these considerations, we have introduced
a new community detection algorithm based on modularity density
maximization. Investigating its performance on Erdős-Rényi and
heterogeneous random networks, we showed that it correctly identifies
them as containing no actual communities. Moreover, our algorithm
outperforms the other existing modularity-density-based method
on every benchmark network that we tested. 
The high level of accuracy it reaches,
its low computational complexity, and the ability to properly
identify networks with no ground-truth communities make it a
powerful tool to investigate complex systems and extract meaningful
information from the network representation of large data sets,
giving it a broad range of application throughout the physical sciences.
At the same time, we have also identified some limitations
of modularity density that were not previously known. More specifically,
we found that the theoretical maximum of modularity density for ring lattices
and pure random trees does not correspond to the trivial partition, but
rather to partitions with more than one community. We find this
particularly intriguing, since Erdős-Rényi graphs are locally tree-like.
Thus, these results seem to suggest a certain relevance of long-distance
links for a correct behaviour of modularity density.
Since most real-world networks are not pure trees or ring lattices,
and indeed do feature shortcut links, we believe these limitations
do not affect the suitability of modularity density and methods based
on it in the analysis and modelling of complex systems.
We will further investigate these limitations in future work.
Additionally, we will also extend this method to other types of
networks, such as bipartite graphs, which require a redefinition
of the concept of community itself.

An implementation of our algorithm is freely available for download at \url{www.fedebotta.com}.

\ack
FB acknowledges the support of UK EPSRC EP/E501311/1.
CIDG acknowledges support by EINS, Network of Excellence
in Internet Science, via the European Commission's FP7
under Communications Networks, Content and Technologies,
grant No.~288021.

\appendix
\section{Implementation details}\label{imple}
Here, we provide a detailed description of the implementation
of the algorithm presented above. To describe how the different
steps are carried out, first we introduce some notation. Let
$P=\left|\mathcal C\right|$ be the size of the current partition.
Then, let $M$ be the partition adjacency matrix of the
network, i.e., the $P\times P$ matrix whose elements $m_{C\widetilde{C}}$
are the number of links between community $C$ and community $\widetilde{C}$.
Also, let $X$ be the community spectra matrix, i.e.,
the $N\times P$ matrix whose elements $x_{iC}$ are the number
of links between node $i$ and nodes in community $C$. Finally,
let $\mathbf S$ be the $P$-dimensional community size vector,
whose elements are the sizes of the communities.

Note that our implementation uses three tolerance parameters:
\begin{enumerate}
\item Power method tolerance $\tau_{pwm}$. This parameter
determines the tolerance for the floating-point comparisons in the
power method.
\item Bisection tolerance $\tau_{bs}$. Since
a bisection with the leading eigenvector of the classical modularity
matrix does not guarantee an increase in modularity density, we introduce
a tolerance $\tau_{bs}$. After each bisection, we check the difference
between the new and old values of modularity density. A bisection is
accepted if modularity density increases or if it decreases by an amount
smaller than $\tau_{bs}$ (more details are given in~\ref{sec:bisection}).
\item Acceptance tolerance $\tau_{acc}$. This parameter defines the
size of the tolerance range when finding the moves that maximally increase
modularity density during tuning and agglomeration steps.
\end{enumerate}

\subsection{Bisection}
\label{sec:bisection}
The first step in the algorithm attempts to bisect a community (see also~\fref{fig2}A),
which can be either the whole network or a previously determined
community, using the traditional modularity matrix. To do so, we
use the spectral method, which we briefly review here. The modularity
matrix $B$ is defined as
\begin{equation*}
B = A_{ij}-\frac{k_i k_j}{2m}\:,
\end{equation*}
and the expression for the modularity of a given partition is
\begin{equation}\label{eq:modularity_matrix}
Q=\frac{1}{2m}\sum_{ij}B_{ij}\delta_{C_iC_j}\:.
\end{equation}
Since we are only considering a potential bisection,
$C_i$ can only assume two values. Thus, a partition
can be represented by a vector $\mathbf s$ whose entries
$s_i$ are $1$ and $-1$ if node $i$ is assigned to the
first or the second community resulting from the split,
respectively. Then, substituting the expression
\begin{equation*}
\delta_{C_iC_j}=\frac{1}{2}(s_i s_j+1)
\end{equation*}
in~\eref{eq:modularity_matrix}, it is
\begin{displaymath}
Q=\frac{1}{4m}\sum_{ij} B_{ij}s_i s_j\:.
\end{displaymath}
The vector $\mathbf s$ can be expressed in terms of the normalized
eigenvectors of $B$ as
\begin{equation*}
\mathbf s=\sum_{i=1}^N \vartheta_i \mathbf{v}_i\:,
\end{equation*}
where the $\vartheta$ are linear combination coefficients,
and $\mathbf{v}_i$ is the $i^\mathrm{th}$ eigenvector of the
modularity matrix, corresponding to the eigenvalue $\lambda_i$.
substituting in~\eref{eq:modularity_matrix}, we obtain
\begin{equation*}
Q=\frac{1}{4m}\sum_{i=1}^N \vartheta_i^2 \lambda_i\:.
\end{equation*}
If we label the eigenvalues so that $\lambda_1>\lambda_2>\cdots >\lambda_N$,
this expression is maximized when $\mathbf s$ is parallel to the leading eigenvector
$\mathbf{v}_1$. However, $\mathbf s$ is a vector whose entries can only be ±1.
Thus, we can only choose its elements to make it as parallel to $\mathbf{v}_1$
as possible. One way of achieving this is to set $s_i=1$ if ${v_1}_i>0$ and
$s_i=-1$ if ${v_1}_i<0$. Then, the bisection consists in finding the leading
eigenvector of $B$ and, if the corresponding eigenvalue is positive,
dividing the nodes according to this rule. Several metohds can be used to diagonalize
$B$. Since we only need to find a single eigenvector, and this step
only provides a starting guess, we choose to use the power method, which offers
a good tradeoff between speed and accuracy.

Further consideration must be given to the fact that we are performing
a bisection based on the modularity matrix, whereas our aim is to maximize
modularity density. The potential problem is that a bisection based on
modularity might not result in a larger value of modularity density. To
avoid this, we introduce a tolerance parameter $\tau_{bs}$, whose role
is to determine the largest possible decrease in modularity density that
we want to accept when bisecting. In other words, if after the bisection
the modularity density of the new partition has decreased by a value larger
than $\tau_{bs}$, we do not accept the split, and keep the original partition.
We consider only one exception to this rule, namely the first iteration
of the bisection. At the start of the algorithm, all nodes are placed
together and we try to bisect the whole network. At this point,
we accept any bisection in order to allow at least a whole iteration of
the whole algorithm. Indeed, if we didn't accept that, both the tuning
and agglomeration steps could not be executed, thus leaving the network
not partitioned. Note that not partitioning the network could be the correct
answer, but we want to make sure that we have considered other partitions
as well at least once. If not partitioning the network is the best answer,
this will be found by the agglomeration step, that will merge all the communities
together.

\begin{figure}
\centering
\includegraphics[width=0.75\textwidth]{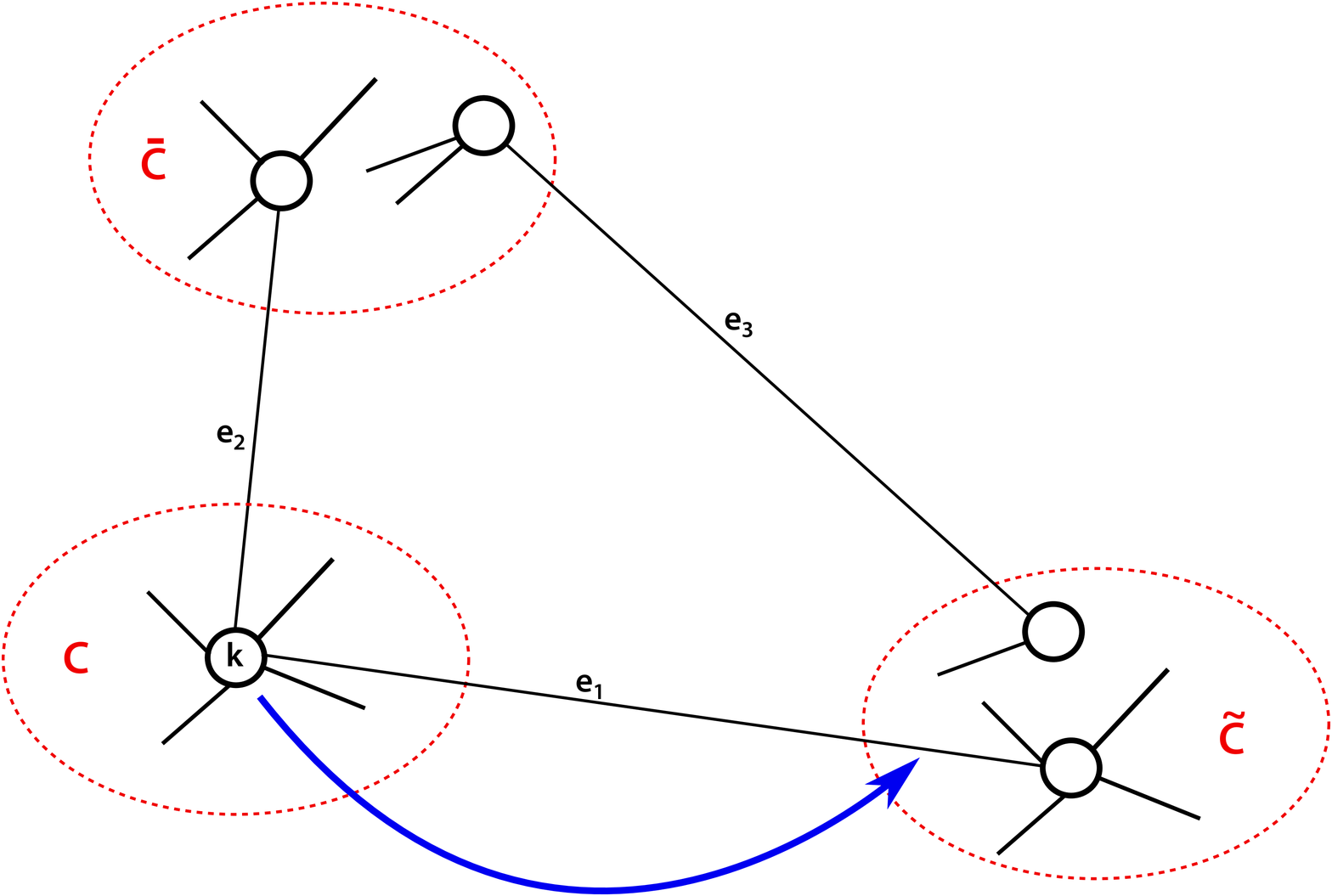}
\caption{Schematic illustration of node $i$ moving from community~$C$ to community~$\widetilde C$.}
\label{fig10}
\end{figure}
Finally, we note that the previous expression for $B$ is correct
only when considering the whole network. When trying to partition a single
community $C$ which does not contain all the nodes, we need to construct
an $n_C\times n_C$ sub-modularity matrix $B^C$ whose elements are
\begin{equation*}
B^C_{ij}=A_{ij}-\frac{k_i k_j}{2m}-\delta_{ij}\left(k_i^C-k_i\frac{k_C}{2m}\right)\:,
\end{equation*}
where $k_i^C$ is the degree of node $i$ within the community
$C$. Using this matrix, we then perform the bisection step
as described above.

In Algorithm~\ref{alg:bisection}, we present a detailed description
of the implementation of this step. For each community, the computation
of the leading eigenvalue through the power method requires $O\left(m_c n_c\right)$
steps. Thus, the worst-case complexity of the the bisection step is
$O\left(mN\right)$.

\begin{algorithm}
\caption{\newline Pseudocode for the bisection step.}
\label{alg:bisection}
\begin{algorithmic}[1]
\Procedure{Bisection Step}{}
	\State flag first bisection $\gets$ 1 \Comment{flag that this is the first bisection}
	\State $w \gets 1$ \Comment{$w$ is the community under consideration}
	\State $\vert S\vert \gets$ 1 \Comment{Current number of communities}
	\While{$w \leq$ $\vert S\vert$} 
		\State current number of nodes $\gets S[w]$ \Comment{$S$ is the community size vector}
		\State current nodes labels $\gets$ find nodes in $S[w]$
		\State $B \gets$ construct $B$ \Comment{$B$ is modularity matrix of the current nodes}
		\If{current number of nodes $>$ 2}
			\State leading $\lambda$, leading $v$ $\gets$ power method(B) 
		\EndIf
		\If{$v$ has at least two negative and two positive components}
			\State flag bisection $\gets 1$
		\EndIf
		\If{$\lambda>0 ~\&$ flag bisection}
			\State bisection($v$, current nodes labels, current number of nodes)
			\State $\vert S\vert \gets \vert S\vert +1$
			\If{old $Q_{ds} -$ new $Q_{ds}>\tau_{bs}$ and flag first bisection$=0$}
				\State cancel bisection
				\State flag$[w]\gets1$
				\State flag fine tuning $\gets$ 0
			\EndIf			
			\If{\small$S[w]>2$ or $S[w+1]>2]$ and flag fine tuning}
				\State fine tuning(current number of nodes, current nodes labels)			
			\EndIf
			\State flag first bisection $\gets$ 0
		\Else
			\State flag$[w] \gets 1$ \Comment{Flag $w$ as blocked}
		\EndIf
		\State flag fine tuning $\gets$ 1
		\If{flag$[w]$}
			\State $w \gets w+1$
		\EndIf
	\EndWhile
\EndProcedure
\end{algorithmic}
\end{algorithm}

\subsection{Tuning Steps}
The crucial part of both the fine tuning and final tuning
steps is that they try to move individual nodes to different
communities (see also~\fref{fig2}B and~C). Thus, we need to consider what happens to the
current partition and how $M$, $X$ and $\mathbf S$
change when we move a node $i$ from community $C$ to community
$\widetilde{C}$. \Fref{fig10} provides an intuitive scheme to
illustrate the changes that follow from such a move. In general,
both the number of internal and external links of $C$ will
change, since node $i$ is leaving this community. However,
to correctly update the modularity density, we also need
to keep track of the changes in all the specific numbers
of links between $C$ and every other community in the current
partition. Similarly, we need to ensure that the internal
and external links of $\widetilde{C}$ are updated correctly.
Finally, the sizes of the two communities changes as well
as a consequence of the move. Below, we describe how to efficiently
perform these updates.

\subsubsection{Updating the Partition Adjacency Matrix}
The partition adjacency matrix $M$ keeps track
of the number of edges between each pair of communities,
as well as the internal number of edges of each community
in its diagonal elements. Looking at~\fref{fig10}, one can
see that the following quantities change:
\begin{itemize}
\item The number of internal links of the community
$C$ that node $i$ is leaving decreases by the internal
degree of node $i$, which is the number of links it
has to other nodes in $C$.
\item The number of internal links of the community
$\widetilde C$ that node $i$ is moving to increases by
the number of links node $i$ has with other nodes in
$\widetilde C$.
\item The number of links between the old and the new community
of node $i$ increases by the number of links between $i$ and its
old community, and decreases by the number of links between $i$
and its new community.
\item The number of links between the old community
$C$ and all the other communities $\bar C\notin\left\lbrace C,\widetilde C\right\rbrace$
decreases by the number of links between $i$ and nodes in $\bar C $.
\item The number of links between the new community
$\widetilde C$ and all the other communities $\bar C\notin\left\lbrace C,\widetilde C\right\rbrace$
increases by the number of links between $i$ and nodes in $\bar C $.
\end{itemize}
In formulae:
\begin{eqnarray*}
m_{C}                    & \to m_{C}-x_{iC} \\
m_{\widetilde{C}}        & \to m_{\widetilde{C}}+x_{i\widetilde{C}} \\
m_{C\widetilde{C}}      & \to m_{C\widetilde{C}} + x_{i C} - x_{i\widetilde{C}} \\
m_{C \bar{C}}           & \to m_{C \bar{C}}-x_{i\bar{C}} \quad\quad\quad\quad\forall\bar C\notin\left\lbrace C,\widetilde C\right\rbrace \\
m_{\widetilde C\bar{C}} & \to m_{\widetilde C\bar{C}} + x_{i\bar{C}} \quad\quad\quad\quad\forall\bar C\notin\left\lbrace C,\widetilde C\right\rbrace\:,
\end{eqnarray*}
where we dropped the repeated index for the diagonal
elements of $M$ to keep the notation consistent.

\subsubsection{Updating the Community Spectra Matrix}
The rows of the matrix $X$ are the community
spectra of the nodes, containing the numbers of links
that each node forms with nodes in all the individual
communities in the current partition. When a node $i$
changes community, its community spectrum does not change.
However, every neighbour of $i$ will experience a change
in the number of connections it has to nodes in the
old and new communities of $i$. In particular, in moving
node $i$ from $C$ to $\widetilde C$, the following changes
happen:
\begin{itemize}
\item Since $i$ is no longer in community $C$, all the nodes
connected to $i$ have one link less to $C$.
\item Since $i$ is now in community $\widetilde C$,
all the nodes connected to $i$ have one connection
more to $\widetilde C$.
\end{itemize}
In formulae:
\begin{eqnarray*}
x_{lC}             & \to x_{lC}-1             & \quad \forall l\mid A_{il}=1 \\
x_{l\widetilde{C}} & \to x_{l\widetilde{C}}+1 & \quad \forall l\mid A_{il}=1\:.
\end{eqnarray*}

\subsubsection{Updating the Community Size Vector}
The updates to this vector are straightforward:
\begin{eqnarray*}
S_C               & \to n_C - 1\\
S_{\widetilde{C}} & \to n_{\widetilde{C}} + 1\:.
\end{eqnarray*}

\subsubsection{Change in Modularity Density}
\begin{algorithm}[t!]
\caption{\newline Pseudocode for the fine tuning step.}
\label{alg:fine}
\begin{algorithmic}[1]
\Procedure{Fine Tuning Step}{}
	\State flag increase $\gets$ 1 \Comment{Flag if there is an increase in modularity density}
	\While{flag increase}
		\State flag increase $\gets$ 0 \Comment{Reset the flag}
		\For{$i_1<$ current number of nodes}
			\For{$i_2<$ current number of nodes}
				\If{flag node$[i_2]=0$}		\Comment{if node $i_2$ is not blocked}
					\If{$x_{i_2,\widetilde{C}}>0$}
						\State $\Delta Q_{ds}[i_2]\gets$ change in $Q_{ds}$ if $i_2$ changes community
					\EndIf
				\EndIf
			\EndFor
			\State $\max \Delta Q_{ds} \gets$ maximum increase in $Q_{ds}$
			\State find all nodes within $\tau_{acc}$ from $\max\Delta Q_{ds}$
			\State node to move $\gets$ pick randomly between nodes with $\max \Delta Q_{ds}$
			\State flag node$[$node to move$]\gets$ 1
			\State fine tuning tree$[i_1]\gets$ fine tuning tree$[i_1-1]+\max \Delta Q_{ds}$
		\EndFor
		\State $\max \Delta Q_{ds}\gets \max($ fine tuning tree$)$
		\If{$\max \Delta Q_{ds} >$ 0}
			\State find all steps within $\tau_{acc}$ of $\max\Delta Q_{ds}$
			\State step in fine tuning tree $\gets$ pick randomly step with $\max \Delta Q_{ds}$
			\State perform all updates in fine tuning tree until the chosen step
			\State flag increase $\gets 1$
		\EndIf
	\EndWhile
\EndProcedure
\end{algorithmic}
\end{algorithm}
Since $Q_{ds}$ is defined as a sum over all current communities,
we consider the terms in its expression~\eref{moddendef} separately,
and show how they change when node $i$ moves from community $C$
to community $\widetilde C$. We first look at what happens to the
contributions of a community $\bar C$ different from $C$ and $\widetilde C$.
In this case, the only changes happen for two terms in the internal
sum:
\begin{equation*}
\fl
\sum_{\hat C\neq\bar C}\frac{m_{\bar C\hat C}^2}{2m n_{\bar C} n_{\hat C}} \to \sum_{\hat C\notin\left\lbrace C,\widetilde C,\bar C\right\rbrace} \frac{m_{\bar C\hat C}^2}{2m n_{\bar C} n_{\hat C}} + \frac{\left(m_{\bar C C} -x_{i\bar C}\right)^2}{2m n_{\bar C}\left(n_C -1\right)} + \frac{\left(m_{\bar C \widetilde C} + x_{i\bar C}\right)^2}{2m n_{\bar C}\left(n_{\widetilde C} +1\right)}\:.
\end{equation*}
Then, we consider the contribution of community $C$:
\begin{eqnarray*}
\fl
\frac{2 m_C^2}{m n_C\left(n_C -1\right)} & \to \frac{2\left(m_C -x_{iC}\right)^2}{m\left(n_C-1\right)\left(n_C-2\right)} \\
\fl
\frac{2 m_C + e_C}{2m} \frac{2m_C}{n_C\left(n_C-1\right)} & \to \frac{2\left(m_C-x_{iC}\right) + e_C + x_{iC}- \sum_{\bar C\neq C}x_{i\bar C}}{2m} \frac{2\left(m_C - x_{iC}\right)}{\left(n_C -1\right)\left(n_C -2\right)}\\
\fl
\sum_{\hat C\neq C} \frac{m_{C \hat C}^2}{2m n_C n_{\hat C}} & \to \sum_{\hat C\notin\left\lbrace C,\widetilde C\right\rbrace} \frac{\left(m_{C\hat C}-x_{i\hat C}\right)^2}{2m\left(n_C -1\right) n_{\hat C} } + \frac{\left(m_{C \widetilde C}+ x_{i C} - x_{i\widetilde C}\right)^2}{2m\left(n_C -1\right) \left(n_{\widetilde C} +1\right) }\:.
\end{eqnarray*}
Finally, we consider the contribution of community $\widetilde C$:
\begin{eqnarray*}
\fl
\frac{2 m_{\widetilde C}^2}{m n_{\widetilde C}\left(n_{\widetilde C} -1\right)} & \to \frac{2\left(m_{\widetilde C} +x_{i\widetilde C}\right)^2}{m\left(n_{\widetilde C} +1\right)n_{\widetilde C}} \\
\fl
\frac{2 m_{\widetilde C} + e_{\widetilde C}}{2m} \frac{2 m_{\widetilde C}}{n_{\widetilde C}\left(n_{\widetilde C} -1\right)} & \to \frac{2\left(m_{\widetilde C} +x_{i\widetilde C}\right) + e_{\widetilde C} -x_{i\widetilde C}+\sum_{\bar C\neq\widetilde C} x_{i\bar C}}{2m} \frac{2\left(m_{\widetilde C} + x_{i\widetilde C}\right)}{\left(n_{\widetilde C} +1\right)n_{\widetilde C}}\\
\fl
\sum_{\hat C\neq\widetilde C} \frac{m_{\widetilde C\hat C}^2}{2m n_{\widetilde C} n_{\hat C} } & \to \sum_{\hat C\notin\left\lbrace C,\widetilde C\right\rbrace}\frac{\left(m_{\widetilde C\hat C} +x_{i\hat C}\right)^2}{2m\left(n_{\widetilde C} +1\right) n_{\hat C}} + \frac{\left(m_{\widetilde CC} + x_{i C} - x_{i\widetilde C}\right)^2}{2m\left(n_{\widetilde C} +1\right)\left(n_C -1\right) }\:.
\end{eqnarray*}

\begin{algorithm}[t]
\caption{\newline Pseudocode for the final tuning step.}
\label{alg:final}
\begin{algorithmic}[1]
\Procedure{Final Tuning Step}{}
	\State flag increase $\gets$ 1 \Comment{Flag if there is an increase in modularity density}
	\While{flag increase}
		\State flag increase $\gets$ 0 \Comment{Reset the flag}
		\For{$i_1<$ N}
			\For{$i_2<$ N}
				\If{flag node$[i_2]=0$} \Comment{if node $i_2$ is not blocked}
				\For{$\bar{C}<\vert S\vert$}
					\If{$x_{i_2,\bar{C}}>0$}	\Comment{if $i_2$ has links to $\bar{C}$}
						\State $\Delta Q_{ds}[i_2][\bar{C}]\gets$ change in $Q_{ds}$ if $i_2$ goes to $\bar{C}$
					\EndIf
				\EndFor
				\EndIf
			\EndFor
			\State $\max \Delta Q_{ds} \gets$ maximum increase in $Q_{ds}$
			\State find all nodes within $\tau_{acc}$ from $\max\Delta Q_{ds}$
			\State node to move $\gets$ pick randomly between nodes with $\max \Delta Q_{ds}$
			\State flag node$[$node to move$]\gets$ 1
			\State final tuning tree$[i_1]\gets$ final tuning tree$[i_1-1]+\max \Delta Q_{ds}$
		\EndFor
		\State $\max \Delta Q_{ds}\gets \max($ final tuning tree$)$
		\If{$\max \Delta Q_{ds} >$ 0}
			\State find all steps within $\tau_{acc}$ of $\max\Delta Q_{ds}$
			\State step in final tuning tree $\gets$ pick randomly step with $\max \Delta Q_{ds}$
			\State perform all updates in final tuning tree until the chosen step
			\State flag increase $\gets 1$
		\EndIf
	\EndWhile
\EndProcedure
\end{algorithmic}
\end{algorithm}
In Algorithm~\ref{alg:fine} and Algorithm~\ref{alg:final},
we present a detailed description of the implementation of
the tuning steps. The complexity of computing the potential
change in modularity density is $O\left(P\right)$, since we
have to consider all the communities to update the split penalty
term. For the fine tuning, this process is repeated $N$ times
per node, yielding a complexity of $O\left(PN^2\right)$. In
the final tuning, instead, all communities are considered
as potential targets, introducing an extra factor of $P$ in
the complexity, which becomes $O\left(P^2N^2\right)$. Note
that these are worst case scenarios, since we typically do
not have to consider all communities for the updates, because
each node is only connected to a subset of them.

\subsection{Agglomeration}
The agglomeration step attempts the merger of pairs of communities (see also~\fref{fig2}D).
If a merger is carried out, a community is obtained whose size is the
sum of the sizes of the original ones. A delicate point is deciding
the label of the new community. In our implementation, we always keep
the smaller of the two labels. So, for instance, if we merge community~1
with community~4, the resulting community will be labelled~1 and community~4
will disappear. We then need to reassign the links of every node in
the network to the new community, and also zero any link to the old
community that disappeared. Below, we describe how to efficiently perform
the required updates, assuming a merger between community~$C$ and community~$\widetilde{C}$
in which the label of the resulting community is~$C$.

\subsubsection{Updating the Partition Adjacency Matrix}
The following changes happen to the partition adjacency matrix:
\begin{itemize}
\item The number of internal links of the merged community
is the sum of the internal links of the two original ones
plus the number of links between the two.
\item All the links of community~$\widetilde{C}$ vanish,
since it has been merged with community $C$.
\item The number of links between the new community
and any other community~$\bar C$ is the sum of the
number of links between each of the two original communities
and~$\bar C$.
\end{itemize}
In formulae:
\begin{eqnarray*}
m_C & \to m_C + m_{\widetilde C}+m_{C\widetilde C} & {}\\
m_{\widetilde C} & \to 0 & {}\\
m_{\widetilde C\bar C} & \to 0 & \quad \forall \bar C\in\mathcal C\\
m_{C\bar C} & \to m_{C\bar C}+m_{\widetilde C\bar C} & \quad \forall \bar C\notin\left\lbrace C,\widetilde C\right\rbrace\:.
\end{eqnarray*}

\subsubsection{Updating the Community Spectra Matrix}
\begin{algorithm}[t!]
\caption{\newline Pseudocode for the agglomeration step.}
\label{alg:agglom}
\begin{algorithmic}[1]
\Procedure{Agglomeration Step}{}
	\State flag increase $\gets$ 1 \Comment{Flag if there is an increase in modularity density}
	\While{flag increase}
		\State flag increase $\gets$ 0 \Comment{Reset the flag}
		\For{$\widetilde{C}<\vert S\vert$}
			\For{$\bar{C}<\vert S\vert$}
				\If{flag community$[\bar{C}]=0$}	\Comment{if $\bar{C}$ is not blocked}
				\For{$\hat{C}<\vert S\vert$}
					\If{flag community$[i_3]=0 ~\& ~m_{\bar{C},\hat{C}}>0$}
						\State $\Delta_{Q_{ds}}[\bar{C}][\hat{C}]\gets$ change in $Q_{ds}$ if we merge $\bar{C}$ and $\hat{C}$
					\EndIf
				\EndFor
				\EndIf
			\EndFor
			\State $\max \Delta Q_{ds} \gets$ maximum increase in $Q_{ds}$
			\State find pairs of communities within $\tau_{acc}$ from $\max\Delta Q_{ds}$
			\State communities to merge $\gets$ pick between those with $\max \Delta Q_{ds}$
			\State flag community$[\bar{C} \vert \hat{C}]\gets$ 1 \Comment{Flag only the one with largest index}
			\State agglomeration tree$[i_1]\gets$ agglomeration tree$[i_1-1]+\max \Delta Q_{ds}$
		\EndFor
		\State $\max \Delta Q_{ds}\gets \max($ agglomeration tree$)$
		\If{$\max \Delta Q_{ds} >$ 0}
			\State step in agglomeration tree $\gets$ picks step with $\max \Delta Q_{ds}$ and smallest number of communities
			\State perform all updates in agglomeration tree until the chosen step
			\State flag increase $\gets 1$
		\EndIf
	\EndWhile
\EndProcedure
\end{algorithmic}
\end{algorithm}
The number of connections between every node $i$ and the merged community
is the sum of the number of links between $i$ and each of the two original
communities, and no node is connected to community~$\widetilde C$ since it
doesn't exist any more:
\begin{eqnarray*}
x_{iC} & \to x_{iC}+x_{i\widetilde C}\\
x_{i\widetilde C} & \to 0\:.
\end{eqnarray*}

\subsubsection{Updating the Community Size Vector}
The changes to the Community Size Vector are once again straightforward:
\begin{eqnarray*}
S_C & \to n_C+n_{\widetilde C}\\
S_{\widetilde C} & \to 0
\end{eqnarray*}

\subsubsection{Change in Modularity Density}
As before, we consider the terms in the definition
of modularity density separately,
showing how they change for the merger considered.
For the contribution of communities $\bar C$ other
than~$C$ and~$\widetilde C$, the only changes happen
in two terms in the internal sum:
\begin{equation*}
\sum_{\hat C\neq\bar C} \frac{m_{\bar C\hat C}^2}{2m n_{\bar C}n_{\hat C}} \to \sum_{\hat C\notin\left\lbrace C,\widetilde C,\bar C\right\rbrace} \frac{m_{\bar C\hat C}^2}{2m n_{\bar C}n_{\hat C}} + \frac{\left(m_{\bar CC} +m_{\bar C\widetilde C}\right)^2}{2mn_{\bar C}\left(n_C+n_{\widetilde C}\right)}\:.
\end{equation*}

\begin{algorithm}[t!]
\caption{\newline Pseudocode for the community detection method.}
\label{alg:commdet}
\begin{algorithmic}[1]
\Procedure{Community Detection}{}
	\State $w\gets 1$ \Comment{Community under consideration}
	\State flag repetition $\gets 1$ \Comment{Flag if there is an increase in modularity density}
	\While{flag repetition}
		\State flag repetition $\gets 0$
		\State $\widetilde{Q}_{ds}\gets Q_{ds}$
		\State {\bf Bisection}
		\If{Current number of communities $>1$}
			\State {\bf Final Tuning}
			\State {\bf Agglomeration}
			\State $\Delta Q_{ds} \gets Q_{ds} - \widetilde{Q}_{ds}$ \Comment{Change in $Q_{ds}$}
			\If{$\Delta Q_{ds}>0$}
				\State $w\gets 1$  \Comment{Restart from the first community}
				\State flag repetition $\gets$ 1 \Comment{Repeat the whole algorithm}
			\EndIf
		\Else
			\State flag repetition $\gets$ 0
		\EndIf
	\EndWhile
\EndProcedure
\end{algorithmic}
\end{algorithm}

Then, we consider the contribution of community~$C$:
\begin{eqnarray*}
\fl
\frac{2m_C^2}{mn_C\left(n_C-1\right)} & \to \frac{2\left(m_C+m_{\widetilde C}+m_{C\widetilde C}\right)^2}{m\left(n_C+n_{\widetilde C}\right)\left(n_C+n_{\widetilde C}-1\right)} \\
\fl
\frac{2m_C+e_C}{2m} \frac{2m_C}{n_C\left(n_C-1\right)} & \to \frac{2\left(m_C+m_{\widetilde C}+m_{C\widetilde C}\right)+e_C+e_{\widetilde C}-2m_{C\widetilde C}}{2m}\\
& \quad\ \times \frac{2\left(m_C+m_{\widetilde C}+m_{C\widetilde C}\right)}{\left(n_C+n_{\widetilde C}\right)\left(n_C+n_{\widetilde C}-1\right)}\\
\fl
\sum_{\bar C\neq C} \frac{m_{C\bar C}^2}{2mn_Cn_{\bar C}} & \to \sum_{\bar C\notin\left\lbrace C,\widetilde C\right\rbrace} \frac{\left(m_{C\bar C}+m_{\widetilde C\bar C}\right)^2}{2m\left(n_C+n_{\widetilde C}\right)n_{\bar C}}\:.
\end{eqnarray*}

Finally, the contribution of community~$\widetilde C$ entirely vanishes.

In Algorithm~\ref{alg:agglom}, we present a detailed description
of the implementation of the agglomeration step. The computational
complexity is $O\left(P^4\right)$. Analogously to the tuning steps,
this is the worst case scenario. In a typical situation, a community
is only connected to a few others, and thus one does not need to
update all the terms in the partition adjacency matrix.

\subsection{Community Detection Algorithm}
Finally, in Algorithm~\ref{alg:commdet} we provide a detailed
description of how the steps presented above are linked together
in our community detection algorithm. The overall complexity of
the algorithm is dominated by the final tuning step, which is the
most computationally expensive, with a complexity $O\left(P^2N^2\right)$.
Along the lines of~\cite{Che13,Che14}, we consider $P$ a constant,
and thus the worst-case complexity reduces to $O\left(N^2\right)$.
To minimize running times, we take advantage of the independence
of the incremental computing steps. Both the fine tuning and final
tuning try to move nodes from one community to a different one.
The calculations of the potential change in modularity density
are independent of each other and thus can be performed in parallel,
rather than serially. This task is fairly straightforward, and
our implementation exploits the widely used C library \emph{Open~MP}
to allow an efficient parallelization using multiple threads on
each computing node during the tuning and agglomeration steps.

\end{document}